# A Possible Generation Mechanism for the IBEX Ribbon from Outside the Heliosphere

S. Grzedzielski, M. Bzowski, A. Czechowski,

Space Research Centre PAS, Bartycka 18A, 00-716 Warsaw, Poland

H.O. Funsten,

Los Alamos National Laboratory, Los Alamos, NM 87545, USA

D.J. McComas,

Southwest Research Institute, San Antonio, TX 78228, USA, and University of Texas, San Antonio, TX 78249, USA

N.A. Schwadron

Boston University, Boston, MA 02215, USA

# ABSTRACT

The brightest and most surprising feature in the first all-sky maps of Energetic Neutral Atoms (ENA) emissions (0.2—6 keV) produced by the Interstellar Boundary Explorer (IBEX) is an almost circular ribbon of a ~140° opening angle, centered at ($l,b$) = (33°, 55°), covering the part of the celestial sphere with the lowest column densities of the Local Interstellar Cloud (LIC). We propose a novel interpretation of the IBEX results based on the idea of ENA produced by charge-exchange between the neutral H atoms at the nearby edge of the LIC and the hot protons of the Local Bubble (LB). These ENAs can reach the Sun's vicinity because of very low column density of the intervening LIC material. We show that a plane-parallel or slightly curved interface layer of contact between the LIC H atoms ($n_H$ = 0.2 cm$^{-3}$, $T$ = 6000—7000 K) and the LB protons ($n_p$ = 0.005 cm$^{-3}$, $T \sim 10^6$ K), together with indirect contribution coming from multiply-scattered ENAs from the LB, may be able to explain both the shape of the ribbon and the observed intensities provided that the edge is < (500—2000) AU away, the LIC proton density is (correspondingly) < (0.04—0.01) cm$^{-3}$, and the LB contains ~1% of non-thermal protons over the IBEX energy range. If this model is correct, then IBEX, for the first time, has imaged in ENAs a celestial object from beyond the confines of the heliosphere and can directly diagnose the plasma conditions in the LB.

*Subject headings:* ISM: bubbles—ISM: clouds—ISM: atoms; ISM: structure—Sun: heliosphere—atomic processes

## 1. Introduction

The first all-sky maps from the Interstellar Boundary Explorer (IBEX) satellite (McComas et al. 2009a) revealed that the most prominent feature of the sky seen in the 0.2-6 keV Energetic Neutral Atom (ENA) emission is a 'ribbon' of enhanced ENA flux (McComas et al., 2009b) forming an almost complete circle in the sky of about 140° diameter and centered on ecliptic/galactic coordinates $(\lambda, \beta)/(l,b) = (221°, 39°)/(33°,55°)$ (Fig.1a) (Funsten et al. 2009a). The ribbon is most pronounced in the 1.1 keV and 1.7 keV energy channels of the IBEX-Hi neutral atom imager (Funsten et al. 2009b), with peak ENA fluxes reaching ~350 and ~120 $(cm^2 \, s \, sr \, keV)^{-1}$, respectively (McComas et al. 2009b). The ribbon is ~20° wide and the average contrast relative to the adjacent slowly varying `distributed flux' at these energies is ~ 2 (Fuselier et al. 2009).

Six possible explanations for this totally unexpected feature in the IBEX observations have been presented (McComas et al. 2009b; Schwadron et al. 2009), one of which, based on ENA emission in proton-neutral atom charge-exchanges (CX) in circumheliospheric plasma, was examined in more detail using a 3-D MHD simulation (Heerikhuisen et al. 2010). The aim of this letter is to suggest a radically different explanation, specifically that ENA emission originates from the Local Bubble (LB), outside of the heliosphere. These ENAs result from protons of the hot, rarified, highly ionized Local Bubble (Breitschwerdt et al. 1998) (~ $10^6$ K, proton and electron density $n_{pLB} = n_{eLB}$ ~0.005 $cm^{-3}$ (Jenkins 2009; Welsh & Shelton 2009)) that neutralize by charge exchange in the interface region with the warm, partly neutral Local Interstellar Cloud (LIC) ($T_{LIC}$ = 6000—7000 K (Lallement & Bertin 1992), neutral H density $n_{H,LIC}$ ~ 0.1—0.2 $cm^{-3}$ (Bzowski et al. 2009), and electron density $n_{eLIC}$ ~ 0.01—0.1 $cm^{-3}$ (Slavin & Frisch 2008)). The contact layer between the LIC and the LB is a natural location for charge exchange, and therefore enhanced ENA emission, in the energy range $\Delta E$ overlapping the range observed by IBEX. The critical questions underlying this explanation of the ribbon are (1) how such a regular, circular feature can emerge, and (2) can the ENA fluxes from this source be sufficiently large to survive the ionization losses during their travel to the inner heliosphere to be measured by IBEX.

It has been inferred (Linsky 1996; Redfield & Linsky 2000; Frisch & Slavin 2006) that the Sun is located very close (perhaps $< 10^4$ AU) to the 'edge' of the LIC in the general direction of the Galactic Center. Thus it may be significant that the brightest arc of the ribbon spans the area of sky with the lowest H column densities of the LIC gas, $N_H < 5 \cdot 10^{16}$ cm$^{-2}$. This is shown in Fig.1b, where the brightest part of the ribbon has been superposed on the map of LIC $N_H$ values from Fig. 5 of Redfield & Linsky (2000). The ribbon of ENA emission may be caused by CX in the particular segment of the LB-LIC interface region that is relatively close to the Sun and least obscured by the LIC material. We show also that a simple interface geometry leads to the observed circular shape.

Most of the ENAs produced at the interface should have LB proton 'thermal' velocities $v_{pLBt}$ corresponding to $\sim 10^2$ eV. However, a turbulent LB should also have a minority population $n_{pLBs}$ of suprathermal protons. This population could produce the ~0.2-6 keV ENAs observed in the ribbon by IBEX and determine both the observed ribbon-to-background contrast and the energy spectra of ENAs.

## 2. The Model

The striking resemblance of the ribbon to emission shells observed in innumerable astrophysical contexts could favor a (quasi-)spherical source, for instance a local protrusion (bay) of the hot plasma (Fig.2b). However, as we show first, a plane-parallel interface (Fig. 2a) can also create a circular ribbon of ENA emissions in the sky.

### 2.1. Plane-parallel interface

In a plane-parallel geometry (Fig. 2a) the ENA intensity $I(\theta)$ (cm$^2$ s sr keV)$^{-1}$ at an angle $\theta$ to the vector $z$ normal to the interface is derived by integrating the omnidirectional ENA emissivity $j_{ENA}(s) = \sigma_{cxs} \, v_{pLBs} \, n_{pLBs} \, n_H / \Delta E$ (cm$^3$ s keV)$^{-1}$ due to the CX reaction H$^+$+H$^0 \rightarrow$ H$^0$+H$^+$, corrected for ENA extinction inside the LB and for obscuration in the LIC, over the line-of-sight $s = z/\cos\theta$. Here $\sigma_{cxs}$, $v_{pLBs}$, $n_{pLBs}$ denote, respectively, the CX cross section, velocity and density of the suprathermal protons in the LB (all functions of energy), and $n_H$ is the density of neutral H in the interface layer, which depends on the details of interface structure (ionization front/shock wave/evaporative boundary, cf. Slavin 1989; Slavin & Frisch 2002).

Current estimates of the properties of the LIC and LB suggest that they can be near pressure equilibrium. The thermal LB pressure is $p_{LB}/k = 2n_{eLB}T_{LB} \sim 10^4$ K cm$^{-3}$ ($k$=Boltzmann constant) assuming the density of the thermal protons is $n_{pLBt} \sim n_{eLB} = 0.005$ cm$^{-3}$ and $T_{LB} = 10^6$ K. Reassessment of the LB conditions, considering that some of the soft X-ray background may come from the solar wind, suggests $p_{LB}/k= 8700$ K/cm$^3$ (Welsh & Shelton 2009). On the other hand the total (thermal + magnetic) LIC pressure $p_{LIC}/k = (n_{HLIC} + 2 \, n_{eLIC}) T_{LIC} + B^2/8\pi k$, for $n_{eLB}$ and $T_{LB}$ as above, is in the range 4200 to 6800 K/cm$^3$. Here $B$ denotes the magnitude of the local interstellar magnetic field, estimated to be 3.1-3.8 µG from the equipartition between the thermal and magnetic energy densities (Frisch & Slavin 2006). An upper limit of 3.8 µG was also inferred from the distances at which the Voyagers crossed the termination shock (Ratkiewicz & Grygorczuk 2008), although fields up to 5.5 µG have recently been suggested (Opher et al. 2009). Based on this, we assume approximate equilibrium between the LIC and LB, with the LIC's edge at a distance $z_0$ from the Sun.

The density $n_H(z)$ of the neutral H evaporating from the LIC follows from the continuity equation

$$\text{div}(n_H \mathbf{v}_H) = -\gamma_{eff} n_H, \qquad (1)$$

where $\gamma_{eff} = \sigma_{cxt} v_{pLBt} n_{pLBt} + \varepsilon_{imp}(T_{LB}) n_{eLB} + \gamma_{ph}$ is the effective ionization rate (s$^{-1}$) of H resulting from CX, electron impact $\varepsilon_{imp}(T_{LB}) n_{eLB}$, and photoionization $\gamma_{ph}$, and $\mathbf{v}_H$ is the average speed of H atoms. In a static case for an isotropic Maxwellian, the velocity component normal to the interface $v_{Hz} = (8 k T_{LIC}/(\pi m_H))^{1/2}$, where $m_H$ is the proton mass. For $T_{LIC} = 6300$ K, $v_{Hz} = 2.9$ km/s. In hydrodynamic solutions (Cowie & McKee 1977) the speed of evaporating medium attains and even exceeds the local sound speed, which changes from ~9 km/s on the LIC side to >100 km/s in the LB at $T_{LB} = 10^6$ K. Therefore we explore $v_{Hz}$ values from ~2.9 to ~25 km/s using $v_{pLBt} = 140$ km/s and, at this speed, $\sigma_{cxt} = 2.9 \cdot 10^{-15}$ cm$^2$ (Lindsay & Stebbings 2005). With $\varepsilon_{imp}(10^6 K) = 5.3 \cdot 10^{-8}$ cm$^3$/s (Olsen et al. 1994), the CX and electron impact ionization rates for H are $2.0 \cdot 10^{-10}$ and $2.7 \cdot 10^{-10}$ s$^{-1}$, respectively. The photoionization rate, estimated using the radiation density from Slavin & Frisch (2008), is $\gamma_{ph} < 10^{-13}$/s and therefore negligible. Assuming at the LB-LIC interface $n_H(z \leq z_0) = n_{H\,LIC}$, Eq. (1) yields $n_H(z) = n_{H\,LIC} \exp[-\gamma_{eff}(z - z_0)/v_{Hz}]$ with $v_{Hz}/\gamma_{eff} = $ ~ 40—360 AU.

With this model the expected ENA intensity $I(\theta)$ results from integration of $j_{ENA}$ over an infinite line-of-sight and accounting for ENA loss by ionization through CX and electron impact inside the LB. This is expressed by:

$$I(\theta) = \frac{1}{4\pi \Delta E} \exp(-\tau_0 \sec\theta) \frac{n_{H\,LIC} n_{pLBs} v_{Hz} v_{pLBs}^2 \sigma_{cxs} \sec\theta}{v_{pLBs} \gamma_{eff} + n_{pLBt} v_{Hz} (\varepsilon_{imp} + v_{pLBs} \sigma_{cxs}) \sec\theta}, \qquad (2)$$

where $\exp(-\tau_0 \sec\theta)$ is the dimming of the ENA intensity by the LIC. The LIC extinction thickness of ENAs

$$\tau_0 = \left(\sigma_{cxs}\, n_{pLIC} + \sigma_{A62}\, n_{He^+LIC} + \sigma_{E2}\, n_{HLIC} + \sigma_{E10}\, n_{HeLIC}\right) z_0 \sim \sigma_{cxs}\, n_{pLIC}\, z_0 \qquad (3)$$

arises primarily from CX of ENAs by LIC protons and secondarily from CX with $He^+$ ($\sigma_{A62}$ = 4.5·10$^{-17}$ cm$^2$) and ionization from interactions with $H^0$ and $He^0$ ($\sigma_{E2}$ = 2.3·10$^{-18}$ cm$^2$ and $\sigma_{E10}$ = 3.6·10$^{-17}$ cm$^2$; both the notation and quoted 1.1 keV values of the cross sections $\sigma_{A62}$, $\sigma_{E2}$, $\sigma_{E10}$ are from Barnett 1990). We use the LIC He I density $n_{HeLIC}$ = 0.015 cm$^{-3}$ derived by Ulysses (Witte et al. 2004). In calculating $\tau_0$, $n_{He^+LIC}$ was chosen to always satisfy the abundance condition H/He = 10 for any value of the unknown $n_{pLIC}$.

From Eq.(2), $I(\theta)$ attains its maximum value at $\theta < 90°$, resulting in a circular arc like the ribbon. We introduce the density ratio $f$ of suprathermal-to-thermal protons (i.e. $n_{pLBs} = f\, n_{pLBt}$). Assuming $f \ll 1$ and that the suprathermal energy distribution has a ~1 keV effective width and peaks near 1.1 keV, one obtains from Eq.(2) the ENA intensity profiles near Earth shown in Fig. 3a for $f$ ranging from 0.36% to 1.3%. In Fig.3a, the contrast ratio $I_{max}/I(0°)$ of maximum intensity $I_{max}$ to that at 0° is ~ 2—2.6, similar to the observed values (cf. McComas et al. 2009b and Fig.2 in Fuselier et al. 2009). We note that $I(\theta)$ depends linearly on the a priori unknown $f$. Nevertheless, the contrast ratios derived in Fig. 3a could only be obtained for the LIC extinction thickness $\tau_0 < \sim 0.15$—0.2, irrespective of adjusting other parameters including $f$ and in particular of the effective outflow velocity $v_{Hz}$ of the H atoms for which the best fitting value for $v_{Hz}$ is in the range 7-25 km/s. Therefore, even for the assumed very low values of $n_{pLIC}$ = 0.01—0.03 cm$^{-3}$, the distance to the edge of LIC is very small, 260—740 AU. Another feature is that $I_{max}$ lies in the narrow range $\theta \sim 75°$—82° instead of ~70°, and is generally insensitive to variation in individual parameters. The difference between the simulated and measured opening half-angle of the ribbon suggests that the plane-parallel interface may need to be replaced with a curved (spherical section) model.

## 2.2. Curvature of the interface

If the edge of the LIC is curved and convex towards the Sun (e.g. a local bay of the LB – Fig. 2b), then ENA emission is brightest (for $\tau_0$ small enough) when viewing close to tangent to the curved surface. We seek a value of the radius of curvature $R$ such that the maximum of emission occurs at $\theta \approx 70°$, in agreement with the IBEX results; actually we assume $I(\theta)$ tends to 0 at $\theta \to 75°$, thus $R = \sin75°/(1-\sin75°)\, z_0 = 28.3\, z_0$.

Here, the IBEX line-of-sight is defined by angle $\theta$ from the vector between the Sun and the closest point of the LB, which is also the axis of symmetry of the model. The ENA intensity from the LB is given by the integral of $j_{ENA}$ over the distance $s$ along the line-of-sight, corrected for the ENA extinction both inside the LB and LIC:

$$I(\theta) = \frac{1}{4\pi\,\Delta E}\exp\left[-\tau_{LIC}(0, s_{near})\right] \int_{s_{near}}^{s_{far}} j_{ENA}(s)\exp\left[-\tau_{LB}(s_{near}, s)\right] ds \quad (\text{cm}^2\,\text{s sr keV})^{-1}. \qquad (4)$$

Here $s_{near}$ and $s_{far}$, both functions of $\theta$, denote distances to the near and far edge (= infinity in practice) of a spherical bay. $\tau_{LB}(s_{near}, s)$ and $\tau_{LIC}(0, s_{near})$ are the ENA extinction thicknesses over the path from $s$ to $s_{near}$ in the LB and then from $s_{near}$ to the Sun in the LIC, respectively. While $\tau_{LIC}(0, s_{near})$ is directly proportional to $s_{near}$, $\tau_{LB}(s_{near}, s)$ is an integral of the varying CX and electron impact probabilities over the flight path. These come from Eq.(1), which we solve assuming spherical symmetry.

The $I(\theta)$ profiles resulting from Eq. (4) show that a contrast ratio $I_{max}/I(0°)$ of ~2-3 can be achieved for larger extinction thicknesses, up to $\tau_0 \sim 0.5$ (Fig. 3b), for values of $f$ and $v_{Hz}$ similar to the plane-parallel model ($0.375\% \leq f \leq 1.8\%$, 7 km/s $\leq v_{Hz} \leq$ 25 km/s). The resulting shortest distances $z_0$ from the Sun to the LIC edge are correspondingly larger, 390 AU $\leq z_0 \leq$ 1100 AU ($R = 28.3\, z_0 = 11\cdot10^3$—$31\cdot10^3$ AU) for the cases shown in Fig. 3b, which include also an often-quoted value of $n_{pLIC} = 0.04$ cm$^{-3}$ (Izmodenov et al. 2003). The model with a curved interface is not only more adjustable through $R$, but numerically sensitive to the opening angle of the ribbon. For example, a decrease of the half-opening angle by 5° compared with Fig. 3b results in a contrast ratio of 2.2 even at $\tau_0 = 0.5$ and $n_{pLIC} = 0.01$ cm$^{-3}$, which implies $z_0 = 1800$ AU and $R = 29000$ AU. If this idea is correct, IBEX will provide an independent measure of the distance to the boundary of the LIC and of its curvature.

## 3. Discussion

The curved interface model clearly fits the IBEX data better, as it enables larger intensity contrast ratios and larger minimum distances to the edge of the LIC than in the plane-parallel model. The range of values $z_0$ ~500—2000 AU resulting from this interpretation of the enhanced ENA emission in the ribbon is small compared to the spatial scales of the LIC and the LB. The component of the heliospheric velocity (26 km/s) in the direction $(\lambda, \beta) = (221°, 39°)$ of the center of the ribbon 46° from the nose of the heliosphere, is 18 km/s; thus, the Sun could enter the LB in 130—540 years if the interface were stationary relative to the LIC gas. If true, this would also have very significant consequences for our understanding of cosmic rays. We list below other obvious questions arising with our model and sketch some possible ways of dealing with them.

1. In addition to the "primary" ENA flux discussed here we also estimated the contribution of "multiple interaction" ENAs that reach the observer only after one or more cycles of ionization and subsequent re-neutralization somewhere inside the LIC. This contribution is not negligible compared to the primary ENA signal and might explain the presence of observed ENA emission outside the ribbon.

2. Another, likely time-dependent, contribution to the ENAs IBEX observes comes from the heliosheath protons that CX with neutral atoms, as envisioned prior to the discoveries of IBEX [McComas et al., 2009a]. The details of these emissions are unknown: the models do not agree well with the IBEX observations (cf Table 1 in Schwadron et al. 2009).

3. The required fraction $f \leq 2\%$ of suprathermal (~1 keV) protons (~10 times the average proton energy at ~$10^6$ K) constitutes ~ 20% of total plasma pressure. Such conditions seem to prevail in the heliosheath as a result of neutral atoms ionization/pickup/acceleration processes (Voyager data: Richardson & Stone 2009). A turbulent LB, with a neutral abundance estimated to be several percent of the plasma, is likely to be as dynamic as the heliosphere.

4. The generally faint emission of the arc of the ribbon between $(l,b)=(100°,15°)$ and $(l,b)=(200°,70°)$ could be explained by stronger extinction in the LIC in this sector (cf. Fig. 5 of Redfield & Linsky 2000). Fluctuations with smaller angular scales are natural in a dynamic LB with velocity variations ~100 km/s (or more) that must generate large pressure differences and thereby emission variations. The linear scale of the fluctuations at the interface can be estimated by comparing the time scale for heating by adiabatic compression $L/c_s$ ($L$ = linear scale along the ridge of ribbon, $c_s$ = speed of sound in the interface ~ 10—100 km/s) with the ~$10^{10}$ s time scale of interface cooling by electron impact excitation and ionization of the H atoms. The resulting spatial scale $L \sim 700$—$7000$ AU seen (for example) from a distance of $\sim 28\ z_0 \sim 20 \cdot 1000$ AU results in an angular scale of $2°$—$15°$ of possible fluctuations.

5. From the observed ribbon ENA energy spectrum $\sim E^{-\kappa_{ribbon}}$ ($\kappa_{ribbon} \sim 1.7$-$2.4$, Funsten at al., 2009a) one may find, by correcting for energy dependent CX probability and ENA losses, the spectral index $\kappa_{LB} = \kappa_{ribbon} + \Delta\kappa$ of the parent LB proton population. $\Delta\kappa$ depends on $n_{pLIC}$ and $z_0$. For $z_0 > \sim 1000$ AU, $\Delta\kappa \sim$ (a few tenths) i.e. $\kappa_{LB}$ is nearer the 'universal' 3 value.

Thus, we have proposed a completely new idea that might explain the remarkable ribbon of ENA emissions observed by IBEX – an idea that isn't based on the interaction between the heliosphere and the LIC, but between the LIC and the LB.

The SRC PAS authors were supported by the Polish MNiSW grants N-NS-1260-11-09 and N-N203-513038. Work on this study by the US authors was supported by the IBEX Mission as a part of NASA's Explorer Program.

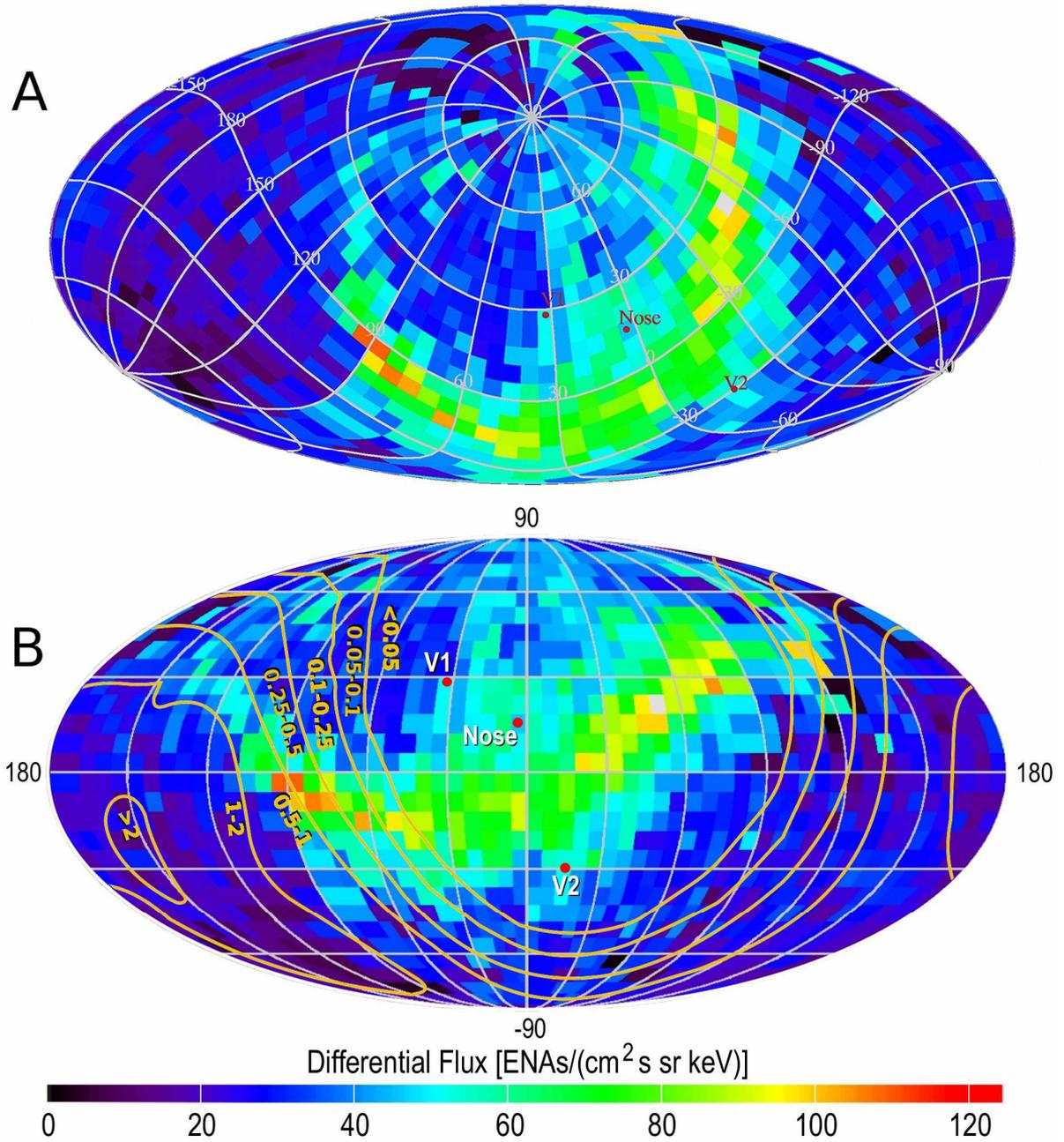

Fig.1. (a) The sky map of the ribbon of ENA emission for the 1.74 keV channel of IBEX-Hi, centered at galactic/ecliptic coordinates $(l,b)/(\lambda,\beta) = (33°, 55°)/(221°, 39°)$, with the grid of galactic coordinates superimposed.

(b) View of the ENA ribbon in galactic coordinates with superposed iso-contours of the LIC neutral H column densities ($N_H$ in $10^{16}/cm^2$) from Redfield and Linsky (2000), which were determined from data on the interstellar absorption to stars well outside the LIC and therefore correspond to upper limits on column densities. "Nose" indicates the direction of motion of the heliosphere in the LIC. "V1" and "V2" show the present positions of the Voyager spacecraft. Color intensity scales identical for (a) and (b).

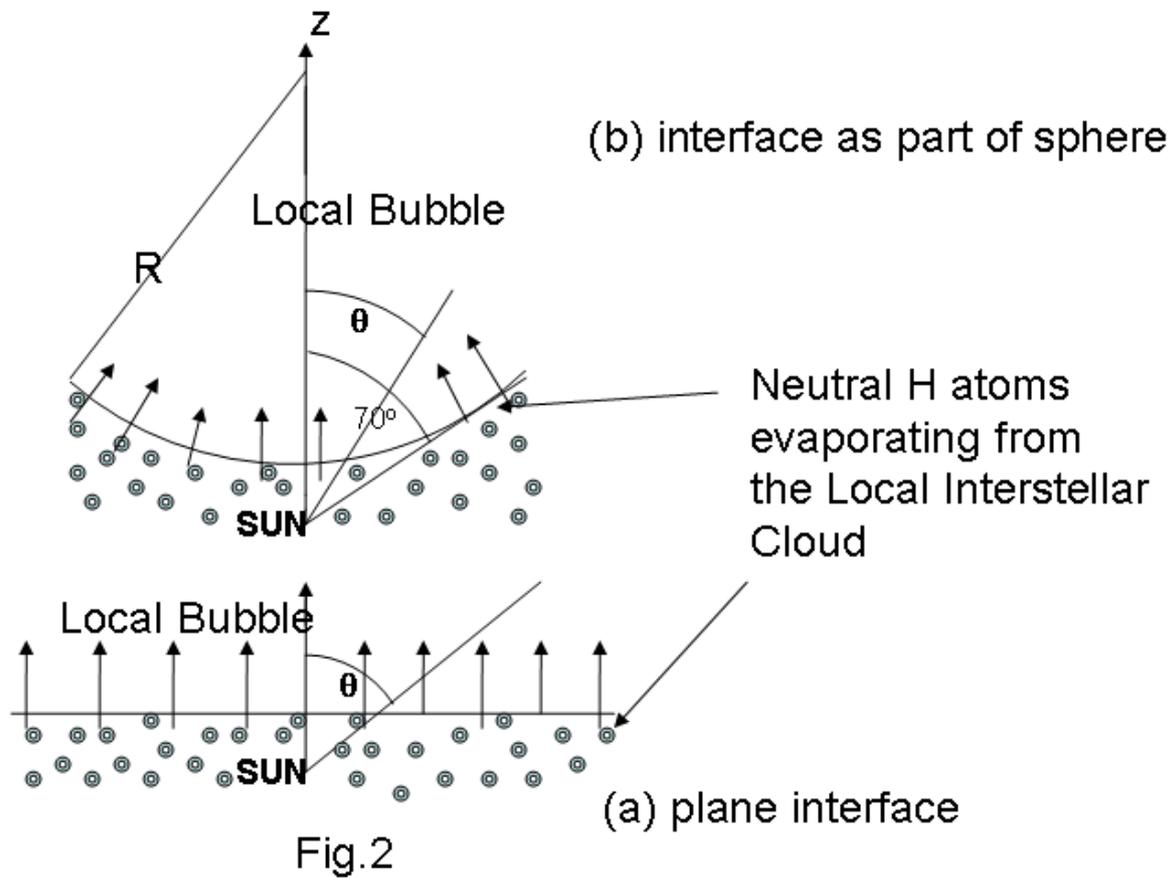

Fig.2. Two models of the interface between the Local Interstellar Cloud and the Local Bubble: (a) plane-parallel layer, (b) a curved layer from a spherical bay of the LB. θ denotes the angle between the IBEX line-of-sight and the shortest distance to the interface.

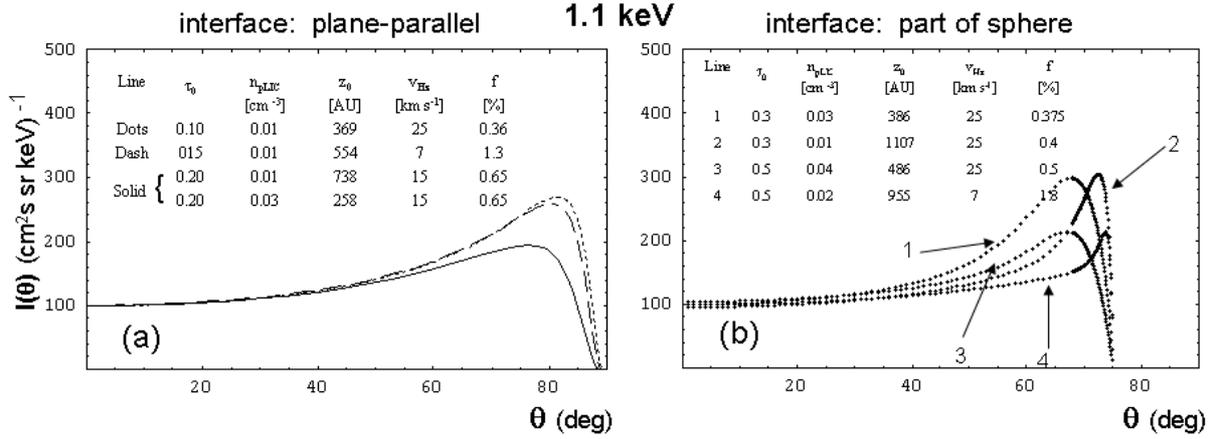

Fig.3. Simulated ENA intensity profiles $I(\theta)$ at 1.1 keV for several assumed extinction thicknesses $\tau_0$ corresponding to shortest flight paths $z_0$ from the interface to the Sun in a uniform LIC. $\theta$ denotes the angle between the line-of-sight and the shortest distance to the interface. (a) plane-parallel interface, (b) curved (spherical bay) interface. To illustrate the influence of various parameters on the most relevant cases of $\tau_0 = 0.20, 0.30, 0.50$, a number of possible combinations of $z_0$ (for given $\tau_0$) and the LIC proton density ($n_{pLIC}$) satisfying Eq. (3) is given. $v_{Hz}$ denotes the assumed evaporation speed of the H atoms and $f$ is the fractional population of suprathermal protons needed in the LB to produce the observed ENA signal.